\documentclass[12pt]{iopart}
\usepackage{latexsym}
\usepackage{amssymb}
\usepackage{amsthm}

\newtheorem{lemma}{Lemma}
\newtheorem{theorem}{Theorem}
\newtheorem{remark}{Remark}
\newcommand{\diag}{\mathop{\mathrm{diag}}\nolimits}
\begin{document}

\title[Detecting community structure with convex optimization]{Identification of community structure in networks with convex optimization}

\author{R Hildebrand}
\address{LJK, Universit\'e Grenoble 1/CNRS, 51 rue des
Math\'ematiques, 38041 Grenoble Cedex 9, France}
\ead{roland.hildebrand@imag.fr}

\begin{abstract}
We reformulate the problem of modularity maximization over the set of partitions of a network as a conic optimization problem over the completely positive cone, converting it from
a combinatorial optimization problem to a convex continuous one. A semidefinite relaxation of this conic program then allows to compute upper bounds on
the maximum modularity of the network. Based on the solution of the corresponding semidefinite program, we design a randomized algorithm
generating partitions of the network with suboptimal modularities. We apply this algorithm to several benchmark networks, demonstrating that it is competitive in accuracy with the best algorithms previously known.
We use our method to provide the first proof of optimality of a partition for a real-world network.
\end{abstract}


\maketitle

\section{Introduction}

A widely accepted measure of community structure in networks is the modularity. The modularity was introduced in \cite{NewmanGirvan04} and is a scalar function defined on the set of partitions of the network.
For a given partition, the modularity describes by how much the intra-community links of the network dominate the links between different communities. Hence the maximizer of the modularity function is the
partition that is best describing the community structure of the network.

However, the problem of maximizing the modularity function over the set of partitions is NP-hard \cite{Brandes_et_al},
and in practice one has to employ algorithms which yield a suboptimal partition. A number of such algorithms
have been proposed by different authors, e.g.\
\cite{NewmanGirvan04},\cite{Radicchi_et_al},\cite{Newman04},\cite{DuchArenas05},\cite{GuimeraSalesAmaral04},\cite{ReichardtBornholdt},\cite{DonettiMunoz},\cite{Newman06}.
A comparison of some available algorithms has been conducted in \cite{Danon_et_al}.
However, none of these algorithms allows to judge the quality of the obtained solution, i.e.\ to tell how close the achieved value of the modularity is to the maximal one.

In this contribution we propose an approach which yields both an upper bound on the maximum of the modularity function, and partitions with suboptimal values of the modularity. We use a formalism developed
by Burer \cite{Burer08} to replace
the discrete search space, namely the set of partitions of the network, by a compact convex set. This set can be described as a compact affine section of the {\it completely positive cone} \cite{HallNewman}.
Thus the combinatorial problem of modularity maximization is replaced by a convex continuous optimization problem,
and the complexity arising from the large number of partitions to be tested translates to the difficulty of deciding membership in a convex set.
Our key idea is to replace the difficult convex set by an
overbounding approximation with an easy description, namely by a {\it semidefinite representable} \cite{NesNem94} set.
Maximizing the modularity over this overbounding approximation amounts to a {\it semidefinite program},
for which numerical solvers are available, and yields an upper bound on the optimal value of the modularity.

We give a simple geometrical interpretation of the employed approximation, which allows to design a randomized method for generating suboptimal partitions from the maximizer of the semidefinite program.
This geometric approach serves also to improve the proposed relaxation further.
These ideas can be considered as a generalization of the semidefinite approach developed by Goemans and Williamson \cite{GoemansWilliamson} for the max-cut problem in combinatorial optimization.

We test the algorithm on several benchmark networks and show that it is among the most accurate methods available in the literature.
The upper bounds on the maximal modularity are in general within a few percent of the achieved suboptimal values. For the Zachary karate club network \cite{Zachary}, the improved version of the 
semidefinite relaxation actually closes the gap between the upper bound and the achieved value of the modularity, thus furnishing an optimality certificate for the obtained partition. To our knowledge,
this is the first proof of optimality for a partition of a real-world network.

The principal drawback of the algorithm is the large computational effort, which limits the range of applicability to networks with a few hundred nodes.

The remainder of the paper is organized as follows. In the next section we provide necessary definitions and background. In Section 3 we reformulate the problem of modularity maximization as an equivalent
convex continuous optimization problem over a compact affine section of the completely positive cone. In Section 4 we derive a semidefinite program yielding upper bounds on the maximal modularity.
In Section 5 we develop a randomized algorithm generating suboptimal partitions of the network into communities.
In Section 6 we test the algorithm on several benchmark networks. In the last section we discuss
the results and line out some suggestions for further improvements.

\section{Definitions and preliminaries}

By $\mathbb R^{k \times l}$ we will denote the space of real $k \times l$ matrices, and by ${\cal S}(n)$ the space of real symmetric $n \times n$ matrices. The space ${\cal S}(n)$ is equipped with an Euclidean
scalar product, the Frobenius inner product $\langle \cdot,\cdot \rangle$ defined by $\langle A,B \rangle = \tr(AB)$.
By ${\bf 1}_k$ we denote a column vector of length $k$ consisting of 1's, and by ${\bf 1}_{k \times l} = {\bf 1}_k{\bf 1}_l^T$ a $k \times l$ matrix consisting of 1's. For a matrix $A$, $vec(A)$ will denote the vector
obtained by stacking the columns of $A$. For matrices $A,B$, $A \otimes B$ will denote the Kronecker product of the matrices $A,B$, i.e.\ a matrix which is obtained from $A$ by replacing each element $A_{ij}$ by
the product $A_{ij}B$. By $I_n$ we denote the $n \times n$ identity matrix. For an $n \times n$ matrix $A$, denote by $\diag A$ the vector consisting of the diagonal elements of $A$.

A {\it symmetric simplex} in $\mathbb R^n$ is a polytope having $n+1$ vertices, such that each vertex is represented by a unit length vector and the scalar product of each pair of distinct vertices
equals $-\frac1n$.

\subsection{Modularity}

The modularity is a scalar function on the set of partitions of a network and was introduced in \cite{NewmanGirvan04}.
Let $G$ be an undirected graph with $n$ vertices, $m$ edges and adjacency matrix $A$. To any partition of the vertex set into disjoint subsets, called communities, we associate a real number, 
called the modularity, defined by
\[ Q = \frac{1}{2m} \sum_{(i,j):C_i = C_j}(A_{ij}-\frac{k_ik_j}{2m}),
\]
where $C_i$ is the community of vertex $i$ and $k_i$ is the number of edges leaving vertex $i$. Note that $k_i$ is the $i$-th element of the vector $A{\bf 1}_n$. Denote the symmetric $n \times n$ matrix with elements
$\frac{1}{2m}(A_{ij}-\frac{k_ik_j}{2m})$ by $B$. We then have
\begin{equation} \label{B}
B = \frac{1}{4m^2}(A \cdot {\bf 1}_n^TA{\bf 1}_n - A{\bf 1}_n{\bf 1}_n^TA),
\end{equation}
and the vector ${\bf 1}_n$ is in the kernel of $B$.

For a particular partition, the modularity of the partition indicates how much the partition matches the community structure of the graph. To detect the community structure of the graph the
modularity has to be maximized over all partitions.

A partition of the vertex set into at most $p$ subsets will be represented by a $\{0,1\}$-matrix $S$ of size $n \times p$. Here each column corresponds to a
community and each row to a vertex. The element $S_{ij}$ is defined as the indicator function of membership of vertex $i$ in the community $j$.
Then $S{\bf 1}_p = {\bf 1}_n$ and the modularity of the partition is given by $Q = \tr(S^TBS) = \langle B,SS^T \rangle$.
The problem of modularity maximization over all partitions into at most $p$ communities hence becomes
\begin{equation} \label{modularity_max}
\max_{S \in \mathbb R^{n \times p}} \langle B,SS^T \rangle:\quad S_{ij} \in \{0,1\}\ \forall\ i = 1,\dots,n,\ j = 1,\dots,p,\ S{\bf 1}_p = {\bf 1}_n.
\end{equation}

\subsection{Semidefinite programs}

The content of this subsection can be found in many recent standard books on convex optimization, e.g.\ \cite{BGFB1},\cite{NesNem94}.
Any real symmetric matrix $X \in {\cal S}(n)$ can be cosidered as a quadratic form on $\mathbb R^n$, defined by $x \mapsto x^TXx$.
If the quadratic form defined by the matrix $X$ is nonnegative everywhere on $\mathbb R^n$, we shall call the matrix $X$ {\it positive semidefinite} (PSD), and write $X \succeq 0$.
The set of PSD matrices in ${\cal S}(n)$
forms a closed convex cone, which will be denoted by $S_+(n)$. By the Frobenius inner product linear functionals on ${\cal S}(n)$ can be identified
with elements of ${\cal S}(n)$, namely by $F(\cdot) = \langle F,\cdot \rangle$.

A {\it semidefinite program} (SDP) is an optimization problem of the form
\begin{equation} \label{SDP}
\min_X \langle F_0,X \rangle: \quad X \in S_+(n),\ \langle F_i,X \rangle = c_i,\ i = 1,\dots,N,
\end{equation}
where $F_0,\dots,F_N$ are given elements of ${\cal S}(n)$, and $c_1,\dots,c_N$ are given real numbers. Such problems can be solved numerically in polynomial time. A list of SDP solvers can be found
e.g.\ in \cite{WikipediaSDP}. We call $X$ a {\it feasible solution} if it satisfies the constraints $X \succeq 0$, $\langle F_i,X \rangle = c_i$ for all $i$, and an {\it optimal solution} if it in addition minimizes
the objective function $\langle F_0,X \rangle$.

It is not hard to see that if we replace some of the equalities in (\ref{SDP}) by inequalities, we can still bring the problem back to the standard form, although with a larger matrix $X$. Also, by changing
the sign of $F_0$ we can convert maximization problems into minimization problems and vice versa.

\subsection{Completely positive cone}

A matrix $X \in {\cal S}(n)$ is said to be {\it completely positive} (CP) if it can be factorized as $X = CC^T$ such that the factor $C$ has only nonnegative elements. It is not hard to check that the set of
CP matrices forms a convex cone in ${\cal S}(n)$, namely the convex conic hull of all PSD rank 1 matrices with nonnegative elements. A PSD matrix with nonnegative elements is also called
{\it doubly nonnegative matrix}, or DNN matrix. It follows that every CP matrix is DNN.
However, for $n \geq 5$ there exist DNN matrices which are not CP \cite{Diananda},\cite{HallNewman}, and the cone of DNN matrices is an overbounding approximation of the CP cone.

The following fundamental result by Burer \cite{Burer08} allows to transform a quadratic optimization problem like (\ref{modularity_max}) to a {\it completely positive program}, i.e.\
a convex continuous optimization problem involving a constraint of membership in the CP cone. Consider the following optimization problem.
\begin{equation} \label{Burer_probl}
\min_{x \in \mathbb R_+^n} x^TQx + 2c^Tx:\quad Ax = b,\ x_i \in \{0,1\} \ \forall\ i \in I \subset \{1,\dots,n\}.
\end{equation}
Here $Q \in {\cal S}(n)$ and $c \in \mathbb R^n$.
Thus we minimize a quadratic function over the positive orthant subject to linear and binary constraints. The linear constraints are given by some $m \times n$ coefficient matrix $A$ and a right-hand side vector
$b \in \mathbb R^m$.
The binary constraints are imposed on some subset $I$ of elements of the search vector $x$.

{\theorem \label{Burer_theorem} \cite[Theorem 3.1]{Burer08}
Consider optimization problem (\ref{Burer_probl}). Assume that its feasible set is nonempty, that the conditions $x \in \mathbb R^n_+$ and $Ax = b$ together imply $x_i \in [0,1]$ for all $i \in I$,
and that there exists a vector $y \in \mathbb R^m$ such that $A^Ty \in \mathbb R_+^n$, $b^Ty = 1$. Then the optimal value of (\ref{Burer_probl}) is equal to the optimal value of the completely positive program
\[ \fl \min_{X \in {\cal S}(n)} \langle Q,X \rangle + 2c^TXA^Ty:\quad AXA^Ty = b,\ \diag(AXA^T) = \diag(bb^T),\] 
\[ \fl (XA^Ty)_i = X_{ii} \ \forall\ i \in I \subset \{1,\dots,n\},\ y^TAXA^Ty = 1,\ X\ \mbox{completely positive}.
\] }

\section{Reformulation as completely positive program}

In this section we reformulate the problem of modularity maximization as an equivalent completely positive program. It is not hard to verify that problem (\ref{modularity_max}) satisfies the assumptions
of Theorem \ref{Burer_theorem} with $x = vec(S)$, $y = \frac1n {\bf 1}_n$, $A^Ty = \frac1n {\bf 1}_{np}$, $Q = B$, $c = 0$, $I = \{1,\dots,np\}$. By this theorem, (\ref{modularity_max}) is equivalent to
the optimization problem
\[ \max_{X \in {\cal S}(np)} \langle B,\sum_{i=1}^p X_{ii} \rangle,\quad \frac1n \sum_{i,j=1}^p X_{ij} {\bf 1}_n = {\bf 1}_n,\ \sum_{i,j=1}^p \diag X_{ij} = {\bf 1}_n,\] \[ \frac1n X {\bf 1}_{np} = \diag X,\
X\ \mbox{completely positive},\ \frac{1}{n^2} \sum_{i,j=1}^p {\bf 1}_n^T X_{ij} {\bf 1}_n = 1.
\]
Here $X$ is an $np \times np$ matrix consisting of $p \times p$ blocks $X_{ij}$ of size $n \times n$ each. Note that the above completely positive program is invariant with respect to simultaneous permutation
of the row and column indices of the blocks (with the same permutation). In other words, the feasible set of the program and its objective function do not change if one replaces $X$ by the product
$(P \otimes I_n)X(P \otimes I_n)^T$, where $P$ is an arbitrary $p \times p$ permutation matrix. We can then group average the program with respect to the symmetric group $S_p$, i.e.\ impose the additional
condition that $X = (P \otimes I_n)X(P \otimes I_n)^T$ for all permutation matrices $P$. Indeed, for any feasible matrix $X$ of the original program, the group average
$\frac{1}{p!} \sum_{P \in S_p} (P \otimes I_n)X(P \otimes I_n)^T$ will be feasible with the same value of the objective function and satisfy the additional invariance condition. We hence assume that all diagonal blocks
of $X$ equal some matrix $X_D \in {\cal S}(n)$, and all off-diagonal blocks equal some matrix $X_O \in {\cal S}(n)$. This leads to the symmetrized completely positive program
\[ \max_{X_D,X_O \in {\cal S}(n)} \langle B, pX_D \rangle,\ \frac{p}{n} X_S {\bf 1}_n = {\bf 1}_n,\ p \diag X_S = {\bf 1}_n,\ \frac1n X_S {\bf 1}_n = \diag X_D,\] \[ X\ \mbox{is CP},\
\frac{p}{n^2} {\bf 1}_n^T X_S {\bf 1}_n = 1,
\]
where $X_S = X_D + (p-1)X_O$. It is not hard to see that the conditions on $X_S$ imply $pX_S = {\bf 1}_n {\bf 1}_n^T$. By defining $X' = \frac{p^2}{p-1}X_D - \frac{1}{p-1}{\bf 1}_{n \times n}$ the program further simplifies to
\begin{equation} \label{CPP_eq}
\max_{X' \in {\cal S}(n)} \langle \frac{p-1}{p}B, X' \rangle,\ \diag X' = {\bf 1}_n,\ \frac1p {\bf 1}_{np \times np} + (I_p - \frac1p {\bf 1}_{p \times p}) \otimes X' \ \mbox{is CP}.
\end{equation}

We obtain the following result.

{\theorem \label{CPP_eqTh} Consider a graph with $n$ vertices and $m$ edges with adjacency matrix $A$ and let the matrix $B$ be defined by (\ref{B}). For every $p \in \mathbb N$, $p \geq 2$, the
maximum of the modularity over the set of partitions of the vertex set of the graph in at most $p$ subsets is given by the optimal value of the completely positive program (\ref{CPP_eq}). }

\medskip

Recall that every CP matrix is DNN. The condition that $\frac1p {\bf 1}_{np \times np} + (I_p - \frac1p {\bf 1}_{p \times p}) \otimes X'$ has nonnegative elements amounts
to the condition that the elements of $X'$ are contained in the interval $[-\frac{1}{p-1},1]$. The condition $\frac1p {\bf 1}_{np \times np} + (I_p - \frac1p {\bf 1}_{p \times p}) \otimes X' \succeq 0$ is
equivalent to the condition $X' \succeq 0$, as the Kronecker product of PSD matrices is again PSD. Thus by relaxing the condition of membership in the CP cone in (\ref{CPP_eq}) to the condition of membership
in the DNN cone we obtain a semidefinite program with a feasible set that overbounds that of the original completely positive program. The optimal value of this
semidefinite program will hence overbound the maximal value of the modularity. However, we prefer a more intuitive geometrical derivation of this semidefinite program in the next section, because it
allows to lay out a randomized algorithm for generating suboptimal partitions.

\section{Upper bound on the maximal modularity}

In this section we construct a semidefinite program whose solution yields an upper bound on the maximal modularity of a graph. The tools employed for this bear resemblance with the methods
proposed in \cite{GoemansWilliamson} for dealing with the so-called {\it max-cut problem}, and in fact can be viewed as their generalization.

Recall that we represent partitions of the vertex set of the graph into at most $p$ subsets by $\{0,1\}$-matrices $S$ of size $n \times p$.
Each row of $S$ is a standard orthonormal basis vector in $\mathbb R^p$. If the partition assigns the $k$-th vertex to the community $l$, then the $k$-th row $s_k$ of the corresponding matrix $S$ is the
$l$-th basis vector $e_l$. Thus all rows of $S$ lie in the intersection of the unit sphere in $\mathbb R^p$ with the affine subspace given by the linear relation $\langle s,{\bf 1}_p \rangle = 1$. Any row vector $s$ in
this intersection can be represented as a sum $s = \frac{1}{p}{\bf 1}_p + \sqrt{\frac{p-1}{p}}v$, where $v$ is a unit length vector in the $p-1$-dimensional linear subspace given by $\langle v,{\bf 1}_p \rangle = 0$.
In this way we can write the rows of $S$ as sums $s_k = \frac{1}{p}{\bf 1}_p + \sqrt{\frac{p-1}{p}}v_k$, and the matrix $S$ as $\frac{1}{p}{\bf 1}_{n \times p} + \sqrt{\frac{p-1}{p}}V$, where the rows of the matrix $V$
are given by the vectors $v_k$.

Recall that the modularity $Q$ of a partition is given by $\langle B,SS^T \rangle$, where $B$ is a fixed real symmetric $n \times n$ matrix depending on the structure of the graph. We obtain
\[ Q = \langle B,(\frac{1}{p}{\bf 1}_{n \times p} + \sqrt{\frac{p-1}{p}}V)(\frac{1}{p}{\bf 1}_{n \times p} + \sqrt{\frac{p-1}{p}}V)^T \rangle = \langle \frac{p-1}{p}B,VV^T \rangle,
\]
because the vector ${\bf 1}_n$ is in the kernel of $B$. The matrix $VV^T \in S_+(n)$ is the Gram matrix of the vectors $v_k$. Note that $\langle v_k,v_l \rangle = 1$ if the vertices $k,l$ belong to the same community,
and $\langle v_k,v_l \rangle = -\frac{1}{p-1}$ if these vertices belong to different communities. Hence the $v_k$ lie at the vertices of a symmetric simplex in the $p-1$-dimensional linear subspace given by
$\langle v,{\bf 1}_p \rangle = 0$. The assignment of the $v_k$ to the different vertices of the simplex corresponds to the assignment of the vertices of the graph to the different communities.

We obtain the following theorem.

{\theorem \label{semidef_theorem}
Consider a graph with $n$ vertices and $m$ edges with adjacency matrix $A$ and let the matrix $B$ be defined by (\ref{B}). For every $p \in \mathbb N$, $p \geq 2$, the optimal value of the semidefinite
program
\begin{equation} \label{relax}
\fl \max_{X \in {\cal S}(n)} \langle \frac{p-1}{p}B,X \rangle: \quad X \succeq 0,\ \diag X = {\bf 1}_n,\ X_{kl} \geq -\frac{1}{p-1}\ \forall\ 1 \leq k < l \leq n
\end{equation}
is an upper bound on the maximum of the modularity over the set of partitions of the vertex set of the graph in at most $p$ subsets. In particular, the optimal value of the semidefinite
program
\[ \fl \max_{X \in {\cal S}(n)} \langle \frac{n-1}{n}B,X \rangle: \quad X \succeq 0,\ \diag X = {\bf 1}_n,\ X_{kl} \geq -\frac{1}{n-1}\ \forall\ 1 \leq k < l \leq n
\]
is an upper bound on the maximal modularity over the set of all partitions. }

\begin{proof}
Let $S$ be the $n \times p$ $\{0,1\}$-matrix corresponding to the partition realizing the maximum of the modularity. Define a matrix $V$ by $S = \frac{1}{p}{\bf 1}_{n \times p} + \sqrt{\frac{p-1}{p}}V$.
Then $X = VV^T$ is a feasible solution for the SDP (\ref{relax}), and the optimal value of the SDP is overbounding the scalar product $Q = \langle \frac{p-1}{p}B,VV^T \rangle$.
\end{proof}

Note that the condition $X_{kl} \leq 1$ for $k \not= l$ is automatically satisfied by all feasible solutions of (\ref{relax}) due to the conditions $X \succeq 0$ and $\diag X = {\bf 1}_n$.

It has to be stressed that only the optimal value of the semidefinite program (\ref{relax}) yields an upper bound to the considered maximum of the modularity function. An arbitrary feasible solution of (\ref{relax})
has no relation to this maximum. However, the theory of semidefinite programming allows to obtain upper bounds without actually solving the semidefinite program.

{\theorem Assume the conditions of the previous theorem. Let $Y \in {\cal S}(n)$ be a matrix with nonpositive off-diagonal elements satisfying $Y - \frac{p-1}{p}B \succeq 0$. Then the quantity
$\frac{p}{p-1} \tr\,Y - \frac{1}{p-1} {\bf 1}_n^TY{\bf 1}_n$ is an upper bound on the maximum of the modularity over the set of partitions of the vertex set of the graph in at most $p$ subsets. }

\begin{proof}
Assume the notations of the theorem. We will show that the quantity in question is an upper bound on the optimal value of the semidefinite program (\ref{relax}). Let $X$ be a feasible solution to (\ref{relax}).
Since the scalar product of two PSD matrices is nonnegative, we have
\begin{eqnarray*}
0 &\leq \langle X,Y - \frac{p-1}{p}B \rangle = \tr\,Y + 2\sum_{i < j} X_{ij}Y_{ij} - \langle X,\frac{p-1}{p}B \rangle \\ &\leq \tr\,Y + 2\sum_{i < j} (-\frac{1}{p-1})Y_{ij} - \langle X,\frac{p-1}{p}B \rangle \\ &=
\frac{p}{p-1}\tr\,Y - \frac{1}{p-1} {\bf 1}_n^TY{\bf 1}_n - \langle X,\frac{p-1}{p}B \rangle. \qedhere
\end{eqnarray*}
\end{proof}

{\remark To look for the best matrix $Y$ in the previous theorem amounts to the semidefinite program
\[ \min_{Y \in {\cal S}(n)} \langle \frac{p}{p-1} I_n - \frac{1}{p-1} {\bf 1}_{n \times n}, Y \rangle,\quad Y_{ij} \leq 0\ \forall\ i \not=j,\ Y - \frac{p-1}{p}B \succeq 0.
\]
This is the {\it dual} program to (\ref{relax}) and it has the same optimal value \cite{BGFB1}. Every feasible solution of the dual program yields an upper bound on the maximum of the modularity. }

\subsection{Sharpening of the approximation}

In this subsection we propose an improved semidefinite approximation by imposing a generalization of the so-called {\it metric inequalities} \cite{DezaLaurentBook}, which are used to tighten relaxations
for max-cut problems.

Let us consider a symmetric simplex in $\mathbb R^{p-1}$ and a collection of vectors $v_1,\dots,v_n$ lying at the vertices of this simplex.
For any three distinct vectors $v_i,v_j,v_k$ we have three possibilities. Either all three vectors lie at the same vertex, or two vectors lie at the same vertex and the third at another vertex, or all three
vectors lie at distinct vertices. Recall that the scalar product between two vectors is either $-\frac{1}{p-1}$ or 1, depending on whether they lie at distinct vertices or not. It is not hard to see that
among the scalar products $\langle v_i,v_j \rangle$, $\langle v_j,v_k \rangle$, and $\langle v_k,v_i \rangle$ there cannot be exactly two equal to 1. Therefore the inequality
$\langle v_i,v_j \rangle+\langle v_j,v_k \rangle-\langle v_k,v_i \rangle \leq 1$ must hold for every triple of distinct indices $(i,j,k)$.  Therefore we can add the conditions
$X_{ij}+X_{jk}-X_{ki} \leq 1$ to relaxation (\ref{relax}). We obtain the following semidefinite program.
\begin{equation} \label{metric_relax}
\max_{X \in S_+(n)} \langle \frac{p-1}{p}B,X \rangle: \diag X = {\bf 1}_n,\ X_{kl} \geq -\frac{1}{p-1}\ \forall\ 1 \leq k < l \leq n,
\end{equation}
\[ X_{ij}+X_{jk}-X_{ki} \leq 1\ \forall\ i,j,k \in \{1,\dots,n\}. \]
A corresponding sharpened version of Theorem \ref{semidef_theorem} then holds for this relaxation.

\section{Randomized algorithm generating suboptimal partitions}

In this section we utilize the geometrical interpretation of the semidefinite relaxation (\ref{relax}) to devise a randomized algorithm yielding good partitions of the graph.
We will need the following lemma.

{\lemma \label{simplex} Let $p \geq 2$ and let $v_1,\dots,v_p$ be the vertices of a symmetric simplex in $\mathbb R^{p-1}$. Let further $U$ be a random orthogonal transformation of $\mathbb R^{p-1}$
drawn uniformly from the Lie group of orthogonal $(p-1) \times (p-1)$ matrices, and let $w_1,\dots,w_p$ be the images of $v_1,\dots,v_p$ under this transformation. Let $W$ be the $p \times p$ matrix
composed of the scalar products $\langle v_k,w_l \rangle$. Then almost surely, in every row of $W$ there will be a unique maximal element. In case of this event, the indices of the maximal elements
of all rows form the set $\{1,\dots,p\}$. }

\begin{proof}
For $p = 2$ the lemma is evident.

Let $p \geq 3$. Consider two distinct elements of the same row of the matrix $W$, say $W_{kl}$ and $W_{km}$. Equality of these elements amounts to the linear condition $(v_l-v_m)^TUv_k = 0$
on the elements of $U$. If the measure of the set of orthogonal matrices satisfying this condition is nonzero, then all tangent vectors to the orthogonal group at $U$ must lie in the corresponding
linear subspace. This amounts to the condition that $v_k(v_l-v_m)^TU$ is a symmetric matrix, i.e.\ $Uv_k$ and $v_l-v_m$ are collinear. This leads to a contradiction with the condition $(v_l-v_m)^TUv_k = 0$.
This proves the first assertion of the lemma.

Now suppose that every row of $W$ has a unique maximal element, and that two rows $k,l$ share the same index $m$ corresponding to this maximal element. Consider the polyhedral cone
$K = \{ v \,|\, \langle v,w_m-w_{m'} \rangle \geq 0\ \forall\ m' \not= m \}$. It is not hard to check that this cone is the convex conic hull of the vectors $-w_{m'}$, $m' \not= m$, and the
minimal scalar product between two unit length vectors of this cone equals $-\frac{1}{p-1}$. This value is attained if and only if these unit length vectors equal $-w_{m_1},-w_{m_2}$ for some $m_1,m_2 \not= m$.
On the other hand, we have $\langle v_k,w_m-w_{m'} \rangle > 0$, $\langle v_l,w_m-w_{m'} \rangle > 0$ for all indices $m' \not= m$. Hence $v_k,v_l$ lie in the interior of the cone $K$, and the
scalar product of these unit length vectors equals the minimal value $-\frac{1}{p-1}$. This leads to a contradiction.
\end{proof}

Let $X$ be the optimal solution of the semidefinite program (\ref{relax}). As outlined in the previous section, $X$ can be interpreted as the Gram matrix of a collection of vectors $v_k$ encoding
the assignment of the vertices of the graph to different communities. These vectors can be defined as the rows of any factor $F$ of the PSD matrix $X$, i.e.\ any $n \times r$ matrix $F$ such that $X = FF^T$,
where $r$ is the rank of $X$.

Ideally, the rank of $X$ does not exceed $p-1$ and the elements of $X$ take on the values $-\frac{1}{p-1}$ and 1.
In this case, the vectors $v_k$ are vertices of a symmetric simplex in $\mathbb R^{p-1}$. The assignment of the $n$ vectors $v_k$ to the $p$ vertices of the
simplex corresponds to the assignment of the $n$ vertices of the graph to $p$ communities. One can compute this assignment as follows.

Construct the vertex set of an arbitrary symmetric simplex in $\mathbb R^p$, for example by factorizing the $p \times p$ rank $p-1$ matrix $\frac{p}{p-1} I_p - \frac{1}{p-1} {\bf 1}_{p \times p}$ and taking the
rows of the factor. Subject the vertex set to a random orthogonal transformation of $\mathbb R^{p-1}$ drawn uniformly from the Lie group of orthogonal matrices,
to obtain a collection of vectors $w_1,\dots,w_p$. Construct the $n \times p$ matrix $W$ of scalar products
$\langle v_k,w_l \rangle$. For $k = 1,\dots,n$, assign the vertex $k$ of the graph to the community whose index corresponds to the maximal element in the $k$-th row of $W$.

By Lemma \ref{simplex}, this procedure almost surely reproduces the original assignment defined by the vectors $v_k$.
By construction the modularity of the partition obtained this way will equal the optimal value of (\ref{relax}) and will hence be
optimal by Theorem \ref{semidef_theorem}.

However, in general the rank $r$ of the solution matrix $X$ exceeds $p-1$ and there are elements of $X$ which lie in the interior of the interval $[-\frac{1}{p-1},1]$. Therefore the collection of vectors $v_k$
obtained from the factorization of $X$ is a subset of the unit sphere in $\mathbb R^r$ such that the scalar products between distinct vectors obey the inequality $\langle v_k,v_l \rangle \geq -\frac{1}{p-1}$.
Nevertheless, the proximity of two vectors $v_k,v_l$ on the unit sphere can be considered as a measure of the tendency of the vertices $k,l$ of the graph to belong to the same community.
We propose to construct candidate partitions of the graph from the set $\{v_1,\dots,v_n\}$ by the following randomized algorithm, which is an adaptation of above construction.

\bigskip

{\bf Algorithm 1.} Solve the semidefinite program (\ref{relax}), factorize the optimal solution $X = FF^T$ to obtain a factor $F$ of size $n \times r$, where $r$ is the rank of $X$. Define the vectors $v_1,\dots,v_n$
as the rows of $F$. If $p-1 > r$, then append the vectors $v_k$, $k = 1,\dots,n$ with $p-r-1$ zeros. Repeat the steps

1. If $p-1 < r$, then draw a random $(p-1)$-dimensional linear subspace of $\mathbb R^r$ from a uniform distribution and project the vectors $v_k$ on this subspace to obtain vectors $v_k' \in \mathbb R^{p-1}$,
$k = 1,\dots,n$. Otherwise define $v_k' = v_k$.

2. Construct the vertex set of an arbitrary symmetric simplex in $\mathbb R^{p-1}$ and subject it to a random orthogonal transformation of $\mathbb R^{p-1}$ drawn from a uniform distribution,
to obtain a collection of vectors $w_1,\dots,w_p$.

3. Construct the $n \times p$ matrix $W$ of scalar products $\langle v_k',w_l \rangle$.

4. For $k = 1,\dots,n$, assign the vertex $k$ of the graph to the community whose index corresponds to the maximal element in the $k$-th row of $W$.

5. Compute the modularity of the obtained partition.

\smallskip

\noindent until the maximal value of the modularity obtained in the sequence of steps 5 makes no further progress. Output the partition that furnished the maximal value of the modularity.

\bigskip

The stopping rule is formulated somewhat vaguely and can be concretized in many ways. For instance, one can stop if $N \geq N_0$ and the last improvement of the modularity occured more than $\alpha N$
iterations ago, where $N$ is the number of the current iteration and $\alpha \in (0,1)$, $N_0 \in \mathbb N$ are prespecified constants. Note also that the algorithm can output a partition that has strictly
less than $p$ communities, because there can be vertices of the symmetric simplex which no vector $v_k$ was assigned to.

\subsection{Partition in 2 communities}

For $p = 2$ both the algorithm proposed above and the semidefinite program (\ref{relax}) considerably simplify. Namely, the inequality constraints $X_{kl} \geq -1$ in (\ref{relax}) are a consequence of the conditions
$X \succeq 0$ and $\diag X = {\bf 1}_n$. We thus obtain the semidefinite program
\begin{equation} \label{relax2}
\max_{X \in {\cal S}(n)} \langle \frac12 B,X \rangle: \quad X \succeq 0,\ \diag X = {\bf 1}_n.
\end{equation}
Algorithm 1 simplifies to the following algorithm,
because the two vertices of a symmetric simplex in $\mathbb R$ are collinear.

\bigskip

{\bf Algorithm 2.} Solve the semidefinite program (\ref{relax2}), factorize the optimal solution $X = FF^T$ to obtain a factor $F$ of size $n \times r$, where $r$ is the rank of $X$. Define the vectors $v_1,\dots,v_n$
as the rows of $F$. Repeat the steps

1. Draw uniformly a random vector $w$ from the unit sphere in $\mathbb R^r$.

2. For $k = 1,\dots,n$, assign the vertex $k$ of the graph to community 1 or community 2 depending on whether the scalar product $\langle v_k,w \rangle$ has positive or negative sign.

3. Compute the modularity of the obtained partition.

\smallskip

\noindent until the maximal value of the modularity obtained in the sequence of steps 3 makes no further progress. Output the partition that furnished the maximal value of the modularity.

\bigskip

Relaxation (\ref{metric_relax}) can also be improved in the case of two communities. Namely, for any collection $\{v_k\}$ of unit length vectors distributed among the vertices of a symmetric simplex in $\mathbb R$
we have the additional condition $\langle v_i,v_j \rangle+\langle v_j,v_k \rangle+\langle v_k,v_i \rangle \geq -1$, because at least two vectors lie at the same vertex. Therefore, in addition to the conditions
$X_{ij}+X_{jk}-X_{ki} \leq 1$, we can add the conditions $X_{ij}+X_{jk}+X_{ki} \geq -1$ to relaxation (\ref{relax2}).

{\remark For $p = 2$ the modularity maximization problem in its equivalent form (\ref{CPP_eq}) reduces to the maximization of the linear function $\frac12 \langle B, \cdot \rangle$ over the
max-cut polytope \cite{DezaLaurentBook}. The semidefinite relaxation (\ref{relax2}) and Algorithm 2 are standard tools for obtaining upper bounds and suboptimal solutions for this problem and were proposed
in \cite{GoemansWilliamson}. }

\section{Examples}

In this section we compute upper bounds on the maximal modularity and test the algorithms presented in the previous section on several benchmark networks used in the literature.

\subsection{Zachary karate club network}

The Zachary karate club network is a social network with 34 nodes studied in \cite{Zachary}. A split of this network in two communities was observed. In the table below we give upper bounds $Q_{upper}$
and achieved values of the modularity $Q_{subopt}$ together with the number of communities in the best partition for different values of $p$.

\begin{table}
\caption{Upper bounds and achieved values of the modularity for the karate club network.}
\begin{indented}
\item[]\begin{tabular}{@{}llll}
\br
$p$&$Q_{upper}$&$Q_{subopt}$&\#\,comm.\\
\mr
2 & $0.376\,4765$ & $0.371\,7949$ & 2 \\
    3 & $0.420\,4657$ & $0.402\,0381$ & 3 \\
    4 & $0.432\,3106$ & $0.419\,7896$ & 4 \\
    5 & $0.435\,3398$ & & 4 \\
    6 & $0.436\,5051$ & & 4 \\
    7 & $0.437\,0969$ & & 4 \\
    34 & $0.438\,6004$ & & \\
\br
\end{tabular}
\end{indented}
\end{table}

For $p \geq 4$ the algorithm retuned the same partition in 4 communities. The partition for $p = 2$ is the same as the one obtained by Newman \cite{Newman04}, and differs from the observed split
by the assignment of one vertex. The partition obtained for $p = 4$ is given by $\{1,2,3,4,8,12,13,14,18,20,22\}$, $\{5,6,7,11,17\}$, $\{9,10,15,16,19,21,23,27,30,31,33,34\}$, $\{24,25,26,28,29,32\}$.
It has a slightly larger modularity than the best one obtained so far in the literature: Duch and Arenas give a value of $0.4188$ \cite{DuchArenas05}, Newman gives a value of $0.419$ \cite{Newman06}.

The solution of the semidefinite program (\ref{relax}) took approximately 1 sec for each value of $p$. 100 iterations of Algorithm 1 were sufficient to obtain the solution for each value of $p$,
which took a fraction of a second of CPU time.

For the karate club network we ran also the sharpened semidefinite program (\ref{metric_relax}) for the values $p = 2,3,34$. For $p = 2$, the inequalities $X_{ij}+X_{jk}+X_{ki} \geq -1$ were added.
For $p = 3$ the upper bound improved to $0.404\,6578$.
For $p = 34$, the optimal solution to relaxation (\ref{metric_relax}) is the one generated by above partition in 4 communities. 
For $p = 2$, the optimal solution is the one generated by above partition in 2 communities.
This proves optimality of these partitions. In other words, the partition returned by the algorithm for $p = 2$ is the best one among all partitions in two communities, and the one returned for $p = 4$
is globally optimal among all partitions.

The running time for solving (\ref{metric_relax}) for each value of $p$ was approximately 3 hours, for the augmented relaxation for $p = 2$ it was 5 h 15 min.

\subsection{Dolphin network}

The dolphin network is a social network with 62 nodes studied in \cite{Dolphins}. In the table below we give upper bounds $Q_{upper}$
and achieved values of the modularity $Q_{subopt}$ together with the number of communities in the best partition for different values of $p$.

\begin{table}
\caption{Upper bounds and achieved values of the modularity for the dolphin network.}
\begin{indented}
\item[]\begin{tabular}{@{}llll}
\br
$p$&$Q_{upper}$&$Q_{subopt}$&\#\,comm.\\
\mr
    2 & $0.411\,9486$ & $0.402\,7333$ & 2 \\
    3 & $0.515\,4178$ & $0.494\,1854$ & 3 \\
    4 & $0.545\,1018$ & $0.526\,7988$ & 4 \\
    5 & $0.549\,8893$ & $0.528\,5194$ & 5 \\
    6 & $0.551\,6863$ & & 5 \\
    7 & $0.552\,6765$ & & 5 \\
    62 & $0.555\,2841$ & & \\
\br
\end{tabular}
\end{indented}
\end{table}

For values $p > 5$ the algorithm returned partitions in 5 communities.

The solution of the semidefinite program (\ref{relax}) took approximately 36 sec for each value of $p$. 100 iterations of Algorithm 1 were sufficient to obtain the solution for each value of $p$,
which took a fraction of a second of CPU time.

\subsection{Random graph}

In this subsection we present results on an artificial benchmark problem which was introduced in \cite{NewmanGirvan04} and used in the literature to compare different algorithms for detecting community structure.
We consider a graph with 128 vertices. The vertex set is artificially divided in 4 communities of 32 vertices each. We will refer to this partition as to the canonical partition.
The edges of the graph are generated randomly and independently such that there are $n_{out}$ expected edges
from each vertex to vertices in different canonical communities, and $n_{in}$ expected edges to vertices within the same canonical community.
These numbers are normalized to satisfy $n_{out}+n_{in} = 16$. The probability of presence
of an edge between a given pair of vertices in different communities is hence $p_{out} = n_{out}/96$, and the probability of presence of an edge between a pair of vertices in the same community is
$p_{in} = n_{in}/31$. The higher the value of $n_{out}$, the more difficult it becomes to detect the community structure of the random graph. The performance of an algorithm is measured by the fraction
of vertices that were assigned correctly as defined by the canonical partition, as a function of $n_{out}$.

The expected modularity of the canonical partition equals approximately $3/4 - n_{out}/16$. This value is empirically valid with an error of $5\cdot 10^{-4}$.
The graph becomes completely random for $n_{out} = 1536/127 \approx 12.0945$, namely when $p_{in} = p_{out} = 16/127$.

We tested Algorithm 1 for the values $n_{out} = 6,7,8,9,10$, generating 10 random graphs for each value. For each random graph, we computed an upper bound on the modularity and ran Algorithm 1
for the values $p = 4$ and $p = 5$. For $p = 5$ the algorithm often returned a partition in 4 subsets, especially for lower values of $n_{out}$. If it returned a partition in 5 subsets, then its modularity was
almost always smaller than the best one obtained for $p = 4$. There were three exceptions, in one out of 10 graphs for the values $n_{out} = 7,8,10$ each. In these cases, the vertices in the fifth community were
all considered as incorrectly classified.

For each value of $n_{out}$, we estimated the mean and the standard deviation of the upper bound on the modularity, of the best modularity achieved by Algorithm 1, and of the fraction of correctly classified vertices,
on the basis of the random sample of 10 graphs. Thus the estimated error on the provided values of the mean equals one third of the listed standard deviations.

\begin{table}
\caption{Upper bounds, achieved values of the modularity, fraction of correctly classified vertices for random graphs.}
\begin{indented}
\item[]\begin{tabular}{@{}llll}
\br
$n_{out}$&$Q_{upper}$&$Q_{subopt}$&\%\,correct vertices\\
\mr
    6 & $0.3760\pm0.0104$ & $0.375\pm0.011$ & $99.06\pm0.62$ \\
    7 & $0.3238\pm0.009$ & $0.313\pm0.012$ & $96.8\pm1.58$ \\
    8 & $0.2915\pm0.006$ & $0.251\pm0.015$ & $85\pm5.47$ \\
    9 & $0.2762\pm0.004$ & $0.207\pm0.008$ & $56.02\pm9.08$ \\
    10 & $0.268\pm0.005$ & $0.196\pm0.006$ & $42.27\pm4.09$ \\
\br
\end{tabular}
\end{indented}
\end{table}

Based on the results of the comparative analysis presented in \cite{Danon_et_al}, our algorithm is outperformed only by the simulated annealing algorithm of \cite{GuimeraSalesAmaral04}
for $n_{out} \leq 8$, and is the most accurate for the higher values of $n_{out}$.

The solution of the semidefinite program (\ref{relax}) took approximately 50 min for each random graph and each value of $p$. $10^6$ iterations of Algorithm 1 were performed to obtain the solution for each value of
$p$, which took approximately 1 hour of CPU time.

\subsection{Jazz musicians}

The Jazz musicians network is a social network with 198 nodes studied in \cite{Jazz}. In the table below we give upper bounds $Q_{upper}$
and achieved values of the modularity $Q_{subopt}$ together with the number of communities in the best partition for different values of $p$.

\begin{table}
\caption{Upper bounds and achieved values of the modularity for the Jazz musicians network.}
\begin{indented}
\item[]\begin{tabular}{@{}llll}
\br
$p$&$Q_{upper}$&$Q_{subopt}$&\#\,comm.\\
\mr
2 & $0.354\,1669$ & $0.315\,3506$ & 2 \\
    3 & $0.450\,0446$ & $0.444\,4694$ & 3 \\
    4 & $0.459\,7013$ & $0.445\,1041$ & 4 \\
    5 & $0.461\,4191$ & & 4 \\
    6 & $0.462\,1197$ & & 4 \\
    7 & $0.462\,4950$ & & 4 \\
    198 & $0.463\,6604$ & & \\
\br
\end{tabular}
\end{indented}
\end{table}

The obtained modularity value is slightly smaller than that of the best known partition of this network, namely $0.4452$ reported in \cite{DuchArenas05}.

The solution of the semidefinite program (\ref{relax}) took approximately 11 hours for each value of $p$. $10^6$ iterations of Algorithm 1 were made to obtain the solution for each value of $p$,
which took approximately 1 hour of CPU time.

\section{Conclusions}

In this paper we considered the problem of community detection in networks. A widely accepted approach to this problem is to maximize the modularity function over the set of all partitions of the network.
We studied the problem of modularity maximization from the viewpoint of convex analysis. Our contribution is threefold.

Firstly, we reformulated the discrete problem of modularity maximization over the set of partitions as a convex continuous optimization problem (Theorem \ref{CPP_eqTh}).
This problem appears in the form of a completely positive program, that is the problem of optimizing a linear objective function over an affine section of the cone of completely positive matrices.
This contribution is more of theoretical nature, because efficient algorithms to solve completely positive programs are not available due to their NP-hardness.

Secondly, we relaxed the obtained completely positive program to a semidefinite program, for which efficient means of solution exist. This relaxation was achieved by a standard approach in convex
optimization, namely the replacement of the cone of CP matrices by the overbounding cone of doubly nonnegative matrices. The optimal value of the semidefinite relaxation is an upper bound on the maximal 
achievable modularity. Our approach explicitly includes the possibility to limit the number of allowed communities in the partitions of the network, and so in fact yields
a series of upper bounds indexed by the maximal number of communities. We provided a simple and intuitive geometrical interpretation of the semidefinite relaxation.
These results are formalized in Theorem \ref{semidef_theorem}.

Thirdly, we proposed a randomized algorithm to generate partitions of the network with suboptimal modularity. This algorithm uses the solution of the above-mentioned semidefinite relaxation as a starting point,
so this semidefinite program must be solved in a preliminary step. The algorithm is described in Section 5 (Algorithm 1).

If we consider only divisions of the network in two communities, the problem of modularity maximization becomes equivalent to the well-known problem of optimization of a linear objective function over
the max-cut polytope. In this case both the semidefinite relaxation and the randomized algorithm considerably simplify. In fact, they reduce to the standard semidefinite optimization approach
to the max-cut problem. We provided explicit descriptions of the simplified semidefinite relaxation and randomized algorithm in Subsection 5.1.

\medskip

We tested our approach on a number of standard benchmark problems. The proposed algorithm proved to be among the most accurate algorithms known to date, and in many cases it yields
the best results known so far. For the karate club network, the improved relaxation (\ref{metric_relax}) actually proves optimality of the obtained partition.

A drawback of the algorithm is the high computational effort, and as a consequence of this, its limited range of applicability to networks with a few hundred nodes. However, with the development of
available SDP solvers this range of applicability will increase.

While the cost of solving the semidefinite relaxation is almost independent (with the exception of partitions in at most 2 communities) of the
maximal number of allowed communities, the amount of required iterations of the randomized algorithm increases with this number. This is because the rank of the solution matrix of the semidefinite program
increases more quickly than the number of allowed communities, and more dimensions have to be projected out. But the projection step in the algorithm tends to destroy the information contained in the
solution matrix, and more iterations of the algorithm have to be conducted to compensate for this loss. However, given the relatively small size of the networks that can be treated, the number of considered
communities is also small, and this is not (yet) a critical issue.

It can be observed that the computed upper bounds rapidly converge when increasing the number of allowed communities in the partition. In most cases the bounds are quite tight, with a relative error
in the percent range between the bound and the achieved value of the modularity. For the random network with parameter value $n_{out} = 6$ the precision of the bound lies at astonishing $0.3\%$ and
is an order of magnitude lower than the standard deviation of the maximal modularity itself.

\medskip

There are several ways to improve the semidefinite relaxations (\ref{relax}) and (\ref{relax2}) further, although at a higher computational cost.

One approach is to pursue the path leading to relaxation (\ref{metric_relax}). More inequalities can be obtained by considering subsets
of 4 and more vectors. The resulting optimization problems are characterized by the presence of one matrix inequality constraint and a large number of linear inequality constraints. For this type of problems
special solution methods exist \cite{Fischer_et_al}, which can treat larger problem instances than standard SDP solvers.

Another approach consists in using tighter semidefinite relaxations for the completely positive cone (see e.g.\ \cite{ParriloThesis}).

\medskip

The methods proposed in this contribution also work for graphs with weighted edges. Therefore it is imaginable to combine them with other modularity optimization methods to treat larger networks. One way would be
to first use some hierachical clustering scheme to reduce the size of the network to several hundred nodes, and then to apply the algorithms proposed here to perform the final optimization.

\ack

The semidefinite programs were solved with the SDPA online solver \\
({\tt http://sdpa.indsys.chuo-u.ac.jp/portal/})
provided by the SDPA project. The author would like to thank the members of this project for this possibility.

This paper presents research results
of the project CARESSE of the pole MSTIC of Universit\'e Joseph Fourier, France.

\end{document}